\documentclass[aps,pra,superscriptaddress,showpacs,twocolumn]{revtex4}

\usepackage{graphicx}
\usepackage{amsmath}
\usepackage{textcomp}
\usepackage{dcolumn}
\usepackage{bm}
\usepackage{multirow}

\begin{document}

\title{High-performance single-photon generation with commercial-grade optical fiber}
\author{Christoph S\"oller}
\email{christoph.soeller@mpl.mpg.de}
\affiliation{Max Planck Institute for the Science of Light, G\"unther-Scharowsky-Str.~1, 91058 Erlangen, Germany}
\affiliation{\mbox{Department of Physics, University of Paderborn, Warburger Str.~100, 33098 Paderborn, Germany}}
\author{Offir Cohen}
\affiliation{Department of Physics and Astronomy, University of Delaware, Newark, DE 19716, USA}
\affiliation{Clarendon Laboratory, University of Oxford, Parks Road, Oxford OX1 3PU, UK}
\author{Brian J. Smith}
\affiliation{Clarendon Laboratory, University of Oxford, Parks Road, Oxford OX1 3PU, UK}
\author{Ian A. Walmsley}
\affiliation{Clarendon Laboratory, University of Oxford, Parks Road, Oxford OX1 3PU, UK}
\author{Christine Silberhorn}
\affiliation{Max Planck Institute for the Science of Light, G\"unther-Scharowsky-Str.~1, 91058 Erlangen, Germany}
\affiliation{\mbox{Department of Physics, University of Paderborn, Warburger Str.~100, 33098 Paderborn, Germany}}

\date{11 April 2011}

\begin{abstract}
High-quality quantum sources are of paramount importance for the implementation of quantum technologies. We present here a heralded single-photon source based on commercial-grade polarization-maintaining optical fiber. The heralded photons exhibit a purity of at least 0.84 and an unprecedented heralding efficiency into a single-mode fiber of 85\%. The birefringent phase-matching condition of the underlying four-wave mixing process can be controlled mechanically to optimize the wavelength tuning needed for interfacing multiple sources, as is required for large-scale entanglement generation.
\end{abstract}


\maketitle

Single-photon and photon-pair states are essential ingredients for quantum optical technologies such as quantum cryptography \cite{Gisin2007}, quantum-enhanced measurements \cite{Giovannetti2004}, and quantum-information processing \cite{Walmsley2005}.
Simple and robust sources of photons in pure quantum states are therefore critical if such technologies are to proliferate and to fulfill their enormous potential. The concept of linear optical quantum computing (LOQC), for example, relies on interactions between single photons to build a scalable all-optical quantum computer \cite{Knill2001}. The ideal single-photon source should thus permit easy low-loss integration into single-mode optical networks and reliably emit pure single photons at a high rate. The photons emitted by distinct copies of the source should further be indistinguishable to allow interference. Note that loss minimization is not only crucial for the realization of quantum technologies beyond proof-of-principle experiments, but also for fundamental tests of quantum mechanics, such as closing the detection loophole in the famous Bell tests \cite{Vertesi2010}.

Semiconductor quantum dots \cite{Michler2000,Santori2002,Zwiller2004} and trapped atoms \cite{Kuhn2002,Beugnon2006} or ions \cite{Maunz2007,Barros2009} have emerged as single-photon sources, but must be operated at cryogenic temperatures or in a vacuum. Sources based on impurities in diamond \cite{Gaebel2004} can be operated at room temperature, but they exhibit non-directional emission and the implementation of two sources providing identical pure photons is a major challenge.
This leaves photon-pair sources based on nonlinear media as the most established and reliable alternative. They have been realized for a wide range of wavelengths with different materials \cite{Burnham1970,Fasel2004,Rarity2005,Chen2005} and can usually be operated at room temperature with little experimental overhead. Photon-pair sources can readily be used as sources of single photons by detecting one photon as a herald for the presence of its colleague, which then, in turn, can be further processed. Two common ways to implement such a heralded single-photon source are spontaneous parametric downconversion (SPDC) in nonlinear crystals and spontaneous four-wave mixing (SFWM) in optical fibers. However, to herald single photons of high purity, these sources long had to be used in combination with narrowband filters to minimize spectral entanglement between the photons of a pair. Yet this approach can never lead to a perfectly pure heralded state \cite{Branczyk2010a}.
Only recently has a breakthrough in spectral engineering resulted in the first factorable photon-pair sources \cite{Valencia2007,Mosley2008,Kuzucu2008,Cohen2009,Halder2009,Soller2010,Evans2010,Eckstein2011}, allowing the generation of heralded single photons in intrinsically pure quantum states.

Two methods for achieving spectral factorability of a photon-pair state can be distinguished. The first relies on exact group-velocity matching (GVM) between a broadband pump and either the created signal or idler photons ($v_{p} = v_{s,i}$). This approach has been implemented with SPDC in nonlinear bulk crystal \cite{Mosley2008} and with SFWM in photonic crystal fiber (PCF) \cite{Cohen2009,Halder2009,Soller2010}. However, it can only be realized at fixed wavelengths governed by the respective medium's dispersion properties. The second method requires a group velocity of the pump in between the group velocities of signal and idler ($v_{s,i} < v_p < v_{i,s}$), as well as a pump bandwidth matched to the phase-matching bandwidth of the medium. This approach offers the important possibility of tuning the signal and idler and of controlling their factorability by changing the pump wavelength and bandwidth, respectively.
It has, so far, only been realized with SPDC \cite{Kuzucu2008,Evans2010,Eckstein2011}.

In this manuscript we experimentally demonstrate that this technique can be applied to SFWM and permits the use of a medium with significant advantages compared to previous sources: standard commercial polarization maintaining (PM) fiber. In contrast to other factorable quantum sources, PM fiber emits into a spatial mode perfectly compatible with single-mode optical fibers. It thus allows the generation of high-purity single photons with unprecedented heralding efficiency. Due to the aforementioned GVM mechanism, the source opens the important possibility of wavelength tuning to adjust the generated state to a given application, whereas prior factorable fiber sources have been limited to very specific wavelengths. We further demonstrate that it is possible to optimize the matching of photons emitted from multiple sources by modifying the fiber birefringence with a standard commercial polarization controller. The aforementioned properties make the presented source an ideal new building block for quantum optical applications and a substantial step toward the large-scale implementation of quantum technologies.

The PM fiber we used as a source was the HB800 from Fibercore Limited. We pumped the fiber along the axis with higher refractive index with a modelocked picosecond Ti:sapphire laser (Coherent MIRA 900-D) set to a center wavelength of 715~nm. This led to the generation of signal and idler photons at 618 and 848~nm, respectively, on the axis orthogonal to the pump by spontaneous four-wave mixing. The photons of a pair were thus both emitted in the silicon detector wavelength range ensuring efficient detection. At the same time, the fiber birefringence and consequently the frequency spread between signal and idler was large enough to keep the contamination of the heralded idler photons by Raman scattering of the strong pump beam at a negligible level, as our results will show.

To generate heralded photons of high purity, special care must be taken to prevent entanglement between signal and idler in any degree of freedom. In the presence of entanglement, the heralded photon is projected into a mixed state which severely limits its ability to interfere with other photons and thus, for example, its usefulness for LOQC. Only if the generated pair state is free of entanglement, that is, in a so-called factorable state, can one of the photons be detected as a herald while the other remains in a pure state. As already mentioned, eliminating spectral entanglement is particularly challenging, since it requires a delicate balance between the group velocities of pump, signal, and idler \cite{U'ren2005}.

\begin{figure}
 \centering
 \includegraphics[width=0.48\textwidth]{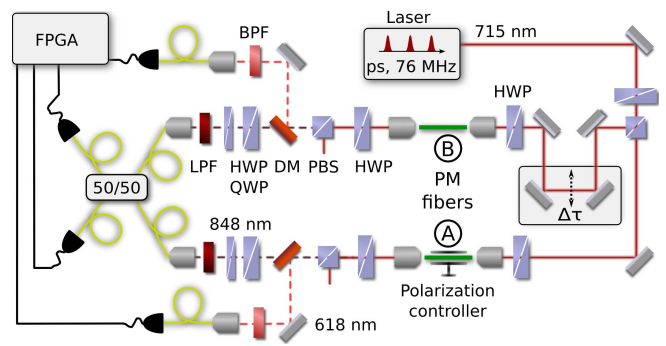}
 \caption{\label{fig:setup} Schematic setup of the HOM interference experiment. The pump beam from our Ti:sapphire oscillator was split and sent to the two PM source fibers. We used half-wave plates (HWP) to align the pump polarization with the high-index axis of each fiber. Polarizing beamsplitters (PBS) at the fiber outputs separated the generated photon pairs from the orthogonally polarized pump. A dichroic mirror (DM) then divided signal and idler before longpass and bandpass filters (LPF, BPF) suppressed residual pump light. The idler photons from both sources interfered at a 50/50 fiber-integrated single-mode beamsplitter with half- and quarter-wave plates (QWP) at the inputs ensuring polarization overlap. The relative time delay ($\Delta \tau$) between the interfering photons could be controlled with a motor-driven delay line in front of source fiber B. Fiber A was placed in a polarization controller to allow fine-tuning of the generated wavelengths. We used silicon single-photon detectors and a field-programmable gate array (FPGA) to record coincidences.}
\end{figure}

To assess the purity of our heralded photons, we conducted a Hong-Ou-Mandel (HOM) interference experiment \cite{Hong1987a} between two separate sources, a test especially important for LOQC.
Figure~\ref{fig:setup} shows our setup which allowed the interference of idler photons coming from the two source fibers while the signal photons served as heralds. Only if the interfering photons are indistinguishable and arrive at the same time, a bunching effect is observed and they always exit the beamsplitter through the same output port. Otherwise, coincidences can be detected due to the two photons exiting through different beamsplitter outputs. The indistinguishability and purity of the heralded single photons can thus be collectively determined by recording the HOM interference dip visibility.

Our pump laser exhibited a secant-squared intensity spectrum whose full-width-at-half-maximum (FWHM) bandwidth could be tuned from approximately 0.2 to 0.5~nm. To achieve factorability of the generated pair state, the pump bandwidth has to be properly matched to the fiber length, which governs the phase-matching width \cite{Smith2009}. The optimum bandwidth to pump our 9~cm source fibers was determined to be 0.33~nm by optimizing the HOM dip visibility. We have already pointed out that this way of achieving factorability by matching medium length and pump bandwidth is different from all previous fiber sources \cite{Halder2009,Cohen2009,Soller2010}. The latter are restricted to fixed signal and idler wavelengths as a result of their GVM technique while our approach opens the possibility of tuning signal and idler. It further leads to the generation of signal and idler photons with equal frequency bandwidths.

\begin{figure*}
   \includegraphics[width=\textwidth]{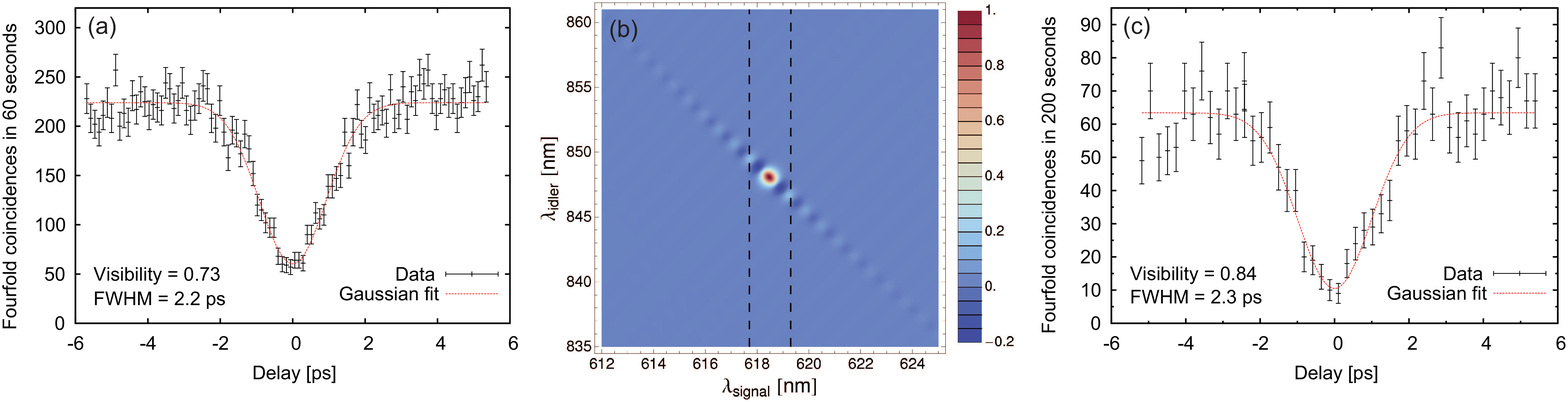}
 \caption{\label{fig:HOM} HOM interference dips. (a) Recorded fourfold coincidences as a function of time delay between the interfering idler photons. Only broadband filters were applied to suppress residual pump light. Error bars have been calculated by taking the square root of each data point, assuming Poissonian count statistics. (b) Illustration of the spectral amplitude distribution of signal and idler. The secondary lobes along the -45\textdegree~axis lead to spectral entanglement and reduce the heralded state's purity. The dashed vertical lines represent the FWHM bandwidth of the filters added for the subsequent measurement. (c) Fourfold coincidences with additional filters in the heralding arms. The heralded state's purity is improved from 0.73 to 0.84 due to suppression of the spectral sidepeaks. The reduced overall count rate can be attributed to limited filter transmission and slightly reduced pump power.}
\end{figure*}

Initial measurements confirmed a purity of the heralded state of at least $0.73\pm0.02$ [see Fig.~\ref{fig:HOM}(a)].
The observed few-picosecond duration of the heralded photons is an important property for potential high data rate, long distance transmission.
The main causes of impurity were the spectral correlations arising from phase matching [see Fig.~\ref{fig:HOM}(b)]. The secondary lobes of the phase-matching function contain only a small fraction of the total spectral intensity, yet restrict the purity to approximately 0.8. In principle, this detrimental effect can be overcome by sophisticated source engineering \cite{U'Ren2006,Branczyk2010}. However, it is also possible to suppress the secondary lobes by filtering without significantly reducing the overall count rate \cite{Smith2009}.

To demonstrate this qualitatively, we set up bandpass filters centered at the signal wavelength in the heralding arms. By adding longpass filters with transmission edges beyond the signal wavelength to the setup and tilting these filters and the broadband filters already in place, we could use this combination of components as a tunable bandpass filter. We set the FWHM bandwidth of these custom filters to approximately 2~nm by monitoring their spectral transmission of a supercontinuum source with a spectrometer (Andor SR-163 with Newton DU-920N).

As expected, the purity of the heralded state improved to $0.84\pm0.04$ [see Fig.~\ref{fig:HOM}(c)]. The observed reduction in overall count rate was mainly due to the limited peak transmission of our customized bandpass filters which led to a decrease by a factor of 5. An additional reduction in count rate by a factor of 2.4 compared to Figure~\ref{fig:HOM}(a) was due to slightly reduced pump power. Calculations have shown that with appropriate filtering the heralded state's purity can be improved further to 0.99, while the reduction in photon flux would essentially be governed by the maximum filter transmission \cite{Smith2009}.

Our recorded purity is comparable to the first recently implemented fiber sources of factorable photon-pair states \cite{Cohen2009,Halder2009}. Note however, that these sources exploited a different, less versatile GVM mechanism and were based on birefringent PCF, a medium with a complicated production process that does not yet offer the same uniformity and quality as standard single-mode fiber.

Photon-pair sources have a small probability of not only emitting one but two or more pairs of photons. To determine the effect of fourfold coincidences caused by these higher photon-number contributions and Raman background on the HOM measurement, we investigated the two sources individually. We measured for each source the rate of threefold coincidences due to clicks in the respective signal and both idler detectors. These events, when coincident clicks in both idler detectors are caused by multi-photon emission from one source, lead to a constant background in the HOM interference measurement. This overall HOM background rate can then be determined by calculating the probability of a threefold event from one source coinciding with a signal detection event from the other source. For the measurements displayed in Figures~\ref{fig:HOM}(a) and (c) this background amounted to 3 counts and 1 count, respectively, in the stated measurement times. The detrimental effect on the recorded HOM visibility is thus, in both cases, approximately 0.01. We further used this recorded threefold coincidence data for each source to calculate the second-order coherence function $g^{(2)}$ for the heralded photons. For the measurement conditions used to record Figure~\ref{fig:HOM}(a) we found $g^{(2)}$ values of $0.012\pm0.001$ and $0.011\pm0.003$ for sources A and B, respectively. The fact that both values are significantly smaller than 1 confirms that the probability of Raman background or higher photon-number contributions contaminating the heralded idler photons is negligible.

An important advantage of our source is its spatial mode profile. Other photon-pair sources based on bulk crystals, nonlinear crystal waveguides, and even PCF do not guide in a spatial mode perfectly matched to standard single-mode fiber. This limits their coupling efficiencies and leads to losses in the collection of the created photons. However, minimizing losses is crucial in to successfully implement quantum technologies on larger scales. Even though remarkably high heralding efficiencies for bulk crystal sources have been achieved \cite{Kurtsiefer2001,Pittman2005,Fedrizzi2007}, this source type lacks the necessary compactness for large-scale applications. At the same time, the few existing sources based on standard fiber \cite{Chen2005} have not yet shown the capability of generating factorable states. To determine our source heralding efficiency we recorded the signal, idler, and coincidence rates for both PM fibers in the setup individually.
Knowing that signal and idler are always emitted as a pair, we can calculate the efficiency $\eta_{h}$ of successfully heralding the presence of an idler photon as $\eta_{h} = R_c / (R_{s} \eta_{d})$. $R_c$ denotes the coincidence rate, $R_{s}$ the signal count rate, and $\eta_d$ corrects for the losses solely caused by the imperfect single-photon detectors. The specified detection probability $\eta_d$ of our single-photon detectors (Perkin Elmer SPCM AQ4C) at 850~nm is ($38\pm1$)\%. After the HOM measurement displayed in Figure~\ref{fig:HOM}(a) we measured 82.6 k signal counts, 177 k idler counts, and 26.6 k coincidences per second for one of our sources. The resulting heralding efficiency for an idler photon is thus ($85\pm1$)\%, a value that agrees well with our previously achieved coupling efficiency during setup alignment. To our knowledge, this is the highest reported heralding efficiency into single-mode fibers for a factorable photon-pair source to date. It illustrates that this source is well suited for single-mode integrated optical networks \cite{Politi2008}. The overall idler detection efficiency, including all optical and detection losses, was as high as ($32.2\pm0.2$)\%. The probability $P_h$ of receiving a heralded photon per pump pulse can be estimated to be $P_h = \eta_h R_s/f_{\text{rep}} = 0.001$, with $f_{\text{rep}} = 76$~MHz.

\begin{table}
\begin{center}
\begin{tabular*}{0.45\textwidth}{@{\extracolsep{\fill}}ccc}
Pressure setting & $\Delta \lambda_{\text{idlerA}}$[nm] & Visibility \\ \hline
1 & \multirow{2}{*}{$\Big\uparrow$} & $0.54 \pm 0.03$ \\
2 &  & $0.70 \pm 0.03$ \\
3 & $0.20 \pm 0.05$ & $0.73 \pm 0.02$ \\
4 & \multirow{2}{*}{$\Big\downarrow$} & $0.69 \pm 0.03$ \\
5 &  & $0.66 \pm 0.04$
\end{tabular*}
\caption{\label{tab:PMscan}Recorded HOM dip visibility for increasing pressure applied to PM fiber A with a polarization controller. The data demonstrate that controlling the fiber birefringence allows optimization of the HOM interference between distinct sources. The increase in idler wavelength for mounting pressure was clearly observable on our spectrometer, confirming the connection between emitted wavelength and fiber birefringence. However, as indicated, the spectrometer resolution was only sufficient to reliably determine the overall wavelength change $\Delta \lambda_{\text{idlerA}} = 0.20\pm0.05$~nm from setting 1 to 5.}
\end{center}
\end{table}

The source tunability is another remarkable feature. It had already been demonstrated that a wide range of signal and idler wavelengths can be accessed by tuning the pump wavelength \cite{Smith2009}. The sensitivity of the generated wavelengths to changes in fiber birefringence can be used to further enhance this tunability and facilitates matching multiple sources.
The frequency spread between signal and idler strongly depends on the amount of fiber birefringence $\Delta n$ \cite{Smith2009}. Placing source fiber A into a polarization controller (Newport F-POL-IL) and applying pressure changes $\Delta n$ and consequently allows fine-tuning of the generated wavelengths.
Since the HOM interference is not only sensitive to the purity of the interfering photons, but also to their spectral overlap, it can be used to detect even small changes in wavelength difference between the two sources. Table~\ref{tab:PMscan} shows the recorded dip visibility for increasing pressure applied to fiber A. At the point of maximum visibility, the photons from both sources are matched best. This feature significantly simplifies interfacing multiple distinct copies of the source.

In conclusion, we have experimentally demonstrated a GVM technique that allows the use of commercial PM fiber as a new source of pure heralded single photons. In addition to the obvious advantages in availability, cost, and ease of use, its high heralding efficiency and precise tunability make this source an ideal building block for quantum optical technologies, where minimizing losses and the ability to interface multiple sources are crucial.

The research leading to these results has received funding from the European Community's Seventh Framework Programme FP7/2007-2013 under Grant No. 213681 through the STREP Project CORNER, under Grant No. 248095 through the Integrated Project Q-ESSENCE and under grant no. CA-ITN-214962-FASTQUAST through the Marie Curie Initial Training Network. It was also supported by the EPSRC through project EP/C51933/0, and by the Royal Society.

\end{document}